\documentclass[twoside,11pt,letter]{article}
\usepackage[margin=1.2in]{geometry}
\usepackage{float}

\usepackage{times}
\usepackage{amsmath}
\usepackage{amsfonts}
\usepackage{amssymb}
\usepackage{graphicx}
\usepackage{mathrsfs}
\usepackage{tabularx}
\usepackage{caption}
\usepackage{bbold}
\usepackage{color}
\usepackage{fancybox}
\usepackage{verbatim}
\usepackage{hyperref}
\usepackage{tikz}
\usetikzlibrary{arrows,automata}
\usetikzlibrary{trees}

\def\bR{\mathbb{R}}

\def\b1{\mathbb{1}}

\def\eE{\mathsf{E}}

\def\tT{\mathtt{T}}

\def\fD{{\mathfrak{D}}}

\def\cV{{\mathcal{V}}}

\def\cov{\mathsf{Cov}}

\def\var{\mathsf{Var}}
\def\e1{\mathsf{1}}

\def\of0{(0)}

\def\d{\partial}

\def\bf0{\mathbf{0}}
\def\cp1{\mathbb{CP}^1}
\def\matn{\mathrm{Mat}_n(\bR)}

\begin{document}

\title{\textbf{Portfolio risk allocation through Shapley value}}
\author{\textbf{Patrick S. Hagan}\\
Gorilla Science\\
PatHagan@GorillaSci.com
\and \textbf{Andrew Lesniewski}\\
Department of Mathematics\\
Baruch College\\
One Bernard Baruch Way\\
New York, NY 10010\\
USA
\and \textbf{Georgios E. Skoufis}\\
Santander Corporate \& Investment Banking\\
2 Triton Square, Regent’s Place\\
London NW1 3AN\\
UK\\
\and \textbf{Diana E. Woodward}\\
Gorilla Science\\
Diana\_Woodward@yahoo.com}

\date{First draft: February14, 2021\\
This draft: \today}
\maketitle

\begin{abstract}
We argue that using the Shapley value of cooperative game theory as the scheme for risk allocation among non-orthogonal risk factors is a natural way of interpreting the contribution made by each of such factors to overall portfolio risk. We discuss a Shapley value scheme for allocating risk to non-orthogonal greeks in a portfolio of derivatives. Such a situation arises, for example, when using a stochastic volatility model to capture option volatility smile. We also show that Shapley value allows for a natural method of interpreting components of enterprise risk measures such as VaR and ES. For all applications discussed, we derive explicit formulas and / or numerical algorithms to calculate the allocations.
\end{abstract}

\titlepage

\section{\label{introSec}Introduction}

All quantitative financial risk management is scenario based. Depending on the the requirements, risk measures are calculated at various granularities: at the single instrument level, portfolio level, enterprise level, or systemic level. In the most myopic case, the risk of a single instrument, one subjects the instrument to specific scenarios in the factors affecting its value. Individual instruments and their risk metrics are aggregated into portfolios, portfolios combine into enterprises, which in turn form an interconnected global web of the financial industry. Moving up the ladder of risk aggregation introduces new categories of risk which require their own scenarios for quantification. The scenarios used for risk quantification may be microscopic (infinitesimal) or macroscopic (finite), they may be generated by a mathematical valuation model or based on historical market prices, they may be defined by the institution carrying the risk or prescribed by a regulating authority.  

A typical situation arises as follows. We consider a portfolio of risky assets whose explainable risk is determined by a number of random state variables $X_1,\ldots,X_n\in\bR$. For mathematical modeling purposes we assume that the random variables $X_i$ have finite first and second moments, $\eE(|X_i|)<\infty$, $\eE(X_i^2)<\infty$, for all $i$. The meaning of the state variables $X_i$ depends on the context. They can represent the variables entering the pricing model of a derivative contract. In the context of portfolio management, one can think of each of the assets in the portfolio as defining a risk factor $X_i$. More generally, we can group subsets of assets in the portfolio into subportfolios, each of which defining a risk factor (such as industry, size, etc). Finally, the value of a financial institution itself may represent a risk factor in a model of systemic risk.

Portfolio risk is measured in terms of a suitable risk metric or a number of risk metrics. One such metric is the portfolio P\&L, i.e. the change in value of the portfolio under a realized or simulated change in the market conditions. One of the key functions of a portfolio manager is to explain (attribute) the daily P\&L to the risk factors $X_i$ of the portfolio. Other metrics, such as value at risk (VaR), expected shortfall (ES), and stress tests are used for economic or regulatory capital calculation, margin sizing, etc. We do not require that these metrics satisfies any (possibly problematic) properties such as coherence. An important question is how to attribute, in a systematic and meaningful way, such a risk metric to each of the risk factors. 

We let $\frak{m}(X)$ denote a generic risk metric, where
\begin{equation}
X=\sum_{j=1}^n\,X_j.
\end{equation}
Risk attribution is straightforward, if the risk factors are orthogonal, i.e. the $X_i$'s are mutually independent. This may be the case if the risk factors are defined in abstract terms, such as principal components, but is typically not true for financially intuitive risk factors. of non-orthogonal, subjecting a portfolio to a particular scenario may partly account for the same risk as it is exhibited in a another, seemingly unrelated scenario. In particular, the risk of the entire portfolio is not simply the sum of its risk measured by all the scenarios.

In this paper we advocate the view that the concept of Shapley value studied in cooperative game theory \cite{MSZ13} offers a ``fair allocation'' design for attributing portfolio risk to each of the risk factors affecting its value. Shapley value allocates overall risk to a risk factor $X_i$ as the average marginal risk of each coalition of risk factors when $X_i$ is added to the coalition. Earlier papers in the context of portfolio management in \cite{MT08}, \cite{O16}, \cite{MBA21}, enterprise risk management \cite{W02}, \cite{BDN20}, and systemic risk \cite{W02}, \cite{TBT10}, \cite{DT13}.

The paper is organized as follows. In Section \ref{shapSec} we review the definition of Shapley value and some of the methods to compute it. Explicit expressions for computing the Shapley value of the variance and volatility games are discussed in Section \ref{varianceSec}. In Section \ref{derSec}, we apply the concept of Shapley value to risk attribution of a portfolio of derivatives with correlated risk factors. Finally, in Section \ref{varSec} we show how Shapley value can help with risk attribution at the enterprise risk management level.

\section{\label{shapSec}Shapley value}

For parts of the discussion in this paper we will find it convenient to use the concepts and language of game theory \cite{MSZ13}. A \textit{coalitional game} is a pair $(\mathscr{N}, v)$ consisting of the \textit{set of players} $\mathscr{N}$ and the \textit{characteristic function} $v:\;2^\mathscr{N}\to\bR$. The only requirement on $v$ is that $v(\emptyset)=0$. Any subset $S\subset\mathscr{N}$ is referred to as a \textit{coalition}. In the context of finance, a risk metric $\frak{m}$ defines a coalitional game whose characteristic function is $v(S)=\frak{m}(X_S)$.

\subsection{Games and risk metrics}

As discussed in the Introduction, a portfolio risk metric $\frak{m}$ is determined by a number of stochastic risk factors $X_j, j=1,\ldots, n$. In the following we will denote by $\mathscr{N}=\{1,\ldots,n\}$ the set of indices labeling the risk factors, and view it as the set of players in a cooperative game. We consider the set of partial risk metrics $\frak{m}(X_S)$, where $S\subset\mathscr{N}$, with the convention that  $\frak{m}(X_\emptyset)=0$. 

The Shapley value \cite{MSZ13} is a way of attributing the value of a coalitional game to each player based on the average marginal value that it contributes to each coalition. Namely, the Shapley value of player $i\in\mathscr{N}$ is defined by
\begin{equation}\label{shdef1}
\mathrm{Sh}_i(v)=\frac{1}{n}\;\sum_{S\subset\mathscr{N}\backslash\{i\}}\,{n-1\choose |S|}^{-1}\big(v(S\cup\{i\})-v(S)\big).
\end{equation}
This can also be recast into the following more symmetric form:
\begin{equation}\label{shdef2}
\mathrm{Sh}_i(v)=\frac{1}{n!}\;\sum_{\pi\in S_n}\,\big(v(S^i_\pi\cup\{i\})-v(S^i_\pi)\big),
\end{equation}
where $S_n$ denotes the set of all permutations of $\mathscr{N}$, and $S^i_\pi=\{j\in\mathscr{N}:\;\pi(j)<i\}$.

A key theorem, due to Lloyd Shapley, states that the Shapley value is the unique way of attributing the value of a coalitional game that satisfies a few natural ``fairness'' requirements \cite{MSZ13}. One of these requirements is the following efficiency property:
\begin{equation}\label{normSh}
\sum_{1=1}^n\,\mathrm{Sh}_i(v)=v(\mathscr{N}),
\end{equation}
stating the sum of individual contributions is equal to the value of the grand coalition.

We conclude this section by deriving an approximate relation between the Shapley values of a game and its square. Assume that the characteristic function $v$ is the square of a characteristic function $\sigma$, i.e. $v(S)=\sigma(S)^2$, for each $S\in 2^{\mathscr{N}}$. Then we can make the approximation
\begin{equation*}
\begin{split}
\mathrm{Sh}_i(\sigma^2)&=\frac{1}{n}\;\sum_{S\subset\mathscr{N}\backslash\{i\}}\,{n-1\choose |S|}^{-1}\big(\sigma(S\cup\{i\})-\sigma(S)\big)\big(\sigma(S\cup\{i\})+\sigma(S)\big)\\
&\approx\mathrm{Sh}_i(\sigma)\lambda,
\end{split}
\end{equation*}
where $\lambda$ represents an average value of $\sigma(S\cup\{i\})+\sigma(S)$. Applying equation \eqref{normSh} to both $\mathrm{Sh}_i(\sigma^2)$ and $\mathrm{Sh}_i(\sigma)$ yields $\lambda\approx\sigma(\mathscr{N})$, and thus
\begin{equation}\label{appShapSqr}
\mathrm{Sh}_i(\sigma)\approx\frac{\mathrm{Sh}_i(\sigma^2)}{\sigma(\mathscr{N})}\,.
\end{equation}

\subsection{\label{mcSec}Calculating Shapley value via Monte Carlo simulations}

Calculating the Shapley value of a game is challenging, owing to the computational complexity of its definition. The only method that is universally applicable is Monte Carlo simulation, first considered in \cite{MS60} in the context of the voting game. Refinements have been discussed in \cite{CGT09} and, for supermodular games, in \cite{LSWW12}.

For the purposes discussed in this paper, especially when the number of players is moderate, the following variation on the original algorithm \cite{MS60} performs sufficiently well. Rather than presenting a pseudocode, below is the actual Python 3 code assuming that the package numpy has been imported as \texttt{np}.
\begin{footnotesize}
\newline\newline\indent
\texttt{def calc\_shapley\_vals(char\_fct, num\_players, sample\_size):
\newline\indent\indent
N = set(np.arange(0, num\_players))
\newline\indent\indent
shapley\_vals = np.zeros(num\_players)
\newline\indent\indent
for i in range(num\_players):
\newline\indent\indent\indent
N\_minus\_i = list(N - \{i\})
\newline\indent\indent\indent
shap = 0.0
\newline\indent\indent\indent
for j in range(sample\_size):
\newline\indent\indent\indent\indent
k = np.random.randint(0, num\_players - i)
\newline\indent\indent\indent\indent
S = np.random.choice(N\_minus\_i, k, replace=False)
\newline\indent\indent\indent\indent
S\_union\_i = np.append(S, i)
\newline\indent\indent\indent\indent
shap += char\_fct(S\_union\_i) - char\_fct(S)
\newline\indent\indent\indent
shapley\_vals[i] = shap / sample\_size
\newline\indent\indent
return shapley\_vals}
\end{footnotesize}
\newline\newline
This code assumes of course that the characteristic function \texttt{char\_fct} has been defined within the appropriate context.

\section{\label{varianceSec}Variance and volatility games}

The variance and volatility games \cite{CSV18} (see also \cite{W02} and \cite{OP16}) are the basis for applicatiosn of Shapley value in portfolio management. In this section we review their properties and derive explicit allocation formulas.

\subsection{\label{anaVarVolSec}Analytic expressions for the variance and volatility games}

Consider $n$ random variables $X_i, i=1,\ldots,n$ distributed so that their first and second moments exist. We consider the \textit{variance game} defined by the following characteristic function. For $S\subset\mathscr{N}$, $v(S)$ is the variance of $X$ conditioned on the complement of $S$,
\begin{equation}
\begin{split}
v(S)&=\var(X|X_{S^c})\\
&=\var(X_S),
\end{split}
\end{equation}
where
\begin{equation}
X_S=\sum_{j\in S}X_j.
\end{equation}
Explicitly,
\begin{equation}
v(S)=\sum_{j,k\in S}\cov(X_j,X_k).
\end{equation}
Note the decomposition:
\begin{equation}\label{covDec}
\begin{split}
v(S\cup T)+v(S\cap T)&=v(S)+v(T)+2\sum_{j\in S\backslash T,\,k\in T\backslash S}\cov(X_j,X_k)\;\\
&=v(S)+v(T)+2\cov(X_{S\backslash T},X_{T\backslash S}),
\end{split}
\end{equation}
for all $S,T\subset\mathscr{N}$.

In particular, if $i\notin S$, then
\begin{equation}
v(S\cup{i})= v(S)+\var(X_i)+2\cov(X_i,X_S).
\end{equation}
Also notice that
\begin{equation*}
\begin{split}
\sum_{\pi\in S_n}\,\cov(X_i,X_{S^i_\pi})&=\sum_{\pi^{-1}\in S_n}\,\cov(X_i,X_{S^i_{\pi^{-1}}})\\
&=\sum_{\pi\in S_n}\,\cov(X_i,X_{\overline S^i_\pi}),
\end{split}
\end{equation*}
where $\overline S^i_\pi=\{j\in\mathscr{N}:\;\pi(j)>i\}$. As a consequence, from \eqref{shdef2},
\begin{equation*}
\begin{split}
\mathrm{Sh}_i(v)&=\frac{1}{n!}\;\sum_{\pi\in S_n}\,\big(\var(X_i)+2\cov(X_i,X_{S^i_\pi})\big)\\
&=\frac{1}{n!}\;\sum_{\pi\in S_n}\,\big(\cov(X_i,X_i+X_{S^i_\pi})+\cov(X_i,X_{\overline S^i_\pi})\big)\\
&=\frac{1}{n!}\;\sum_{\pi\in S_n}\,\cov(X_i,X_i+X_{S^i_\pi}+X_{\overline S^i_\pi})\\
&=\frac{1}{n!}\,\cov\Big(X_i,\sum_{\pi\in S_n}\,X\Big).
\end{split}	
\end{equation*}
This leads to the following equation, established in \cite{CSV18}: 
\begin{equation}\label{varShap}
\mathrm{Sh}_i(v)=\cov(X_i,X).
\end{equation}

We are unaware of a closed form expression for the Shapley value of the \textit{volatility game} with characteristic function given by
\begin{equation}
\sigma(S)=\sqrt{\var(X_S)}.
\end{equation}
However, approximation \eqref{appShapSqr} leads to the following intuitively appealing formula:
\begin{equation*}
	\mathrm{Sh}_i(\sigma)\approx\frac{\cov(X_i,X)}{\sigma(X)}\,,
\end{equation*}
or
\begin{equation}\label{apprSh}
\begin{split}
\mathrm{Sh}_i(\sigma)\approx\rho(X_i,X)\sigma(X_i),
\end{split}
\end{equation}
where $\rho(X_i,X)$ denotes the correlation coefficient between $X_i$ and $X$. 

The attribution of risk factor $X_i$ is proportional to the volatility of that factor, with the proportionality coefficient reflecting the correlation between the factor and the overall risk. Below we verify numerically that this approximation is remarkably accurate.

\subsection{\label{numCovSec}Numerical simulations}

In Table \ref{varianceTab} below we present the results of a test of these results based on simulated data. In particular, we test the accuracy of the analytic approximation \eqref{apprSh} for the volatility game.

The data has been prepared as follows.
\begin{itemize}
\item[(i)] {We generate a $25\times 25$ random (positive definite) covariance matrix $\Sigma$.}
\item[(ii)] {We generate 500 independent samples $X_i$ from the multivariate Gaussian distribution $N(0,\Sigma)$, $X_i\sim N(0,\Sigma)$, $i=1,\ldots, 500$.}
\end{itemize}
We chose 500 observations, as this corresponds to the observation window of 2 years, frequently chosen in the financial industry for VaR calculation.

The meaning of the columns in Table \ref{varianceTab} is as follows:
\begin{itemize}
\item[1.]{SV represents $\widehat{\mathrm{Sh}}_i(v)$,}
\item[2.]{CV represents  $\cov(X_i,X)$,}
\item[3.]{SS represents $\widehat{\mathrm{Sh}}_i(\sigma)$,}
\item[4.]{CS represents  $\rho(X_i,X)\sigma(X_i)$,}
\end{itemize}
for $i=1,\ldots,25$. For the Monte Carlo computations of the Shapley value (columns one and three), we used 100,000 simulations.

\begin{table}[h]
\centering
\begin{tabular}{l |l  || l | l}
SV & CV & SS	& CS \\
\hline\hline
1,268.1	&	1,191.9	&	8.9	&	6.6	\\
1,868.7	&	1,837.0	&	10.6	&	10.2	\\
3,371.2	&	3,355.5	&	17.0	&	18.7	\\
3,992.8	&	4,042.3	&	18.4	&	22.5	\\
1,260.3	&	1,231.1	&	7.9	&	6.9	\\
1,371.3	&	1,310.7	&	8.7	&	7.3	\\
3,018.9	&	3,034.1	&	14.8	&	16.9	\\
4,226.8	&	4,306.9	&	18.7	&	24.0	\\
799.7	&	740.3	&	6.6	&	4.1	\\
3,436.6	&	3,476.4	&	16.2	&	19.4	\\
1,002.0	&	930.3	&	7.5	&	5.2	\\
198.4	&	145.3	&	3.7	&	0.8	\\
693.2	&	686.0	&	4.4	&	3.8	\\
2,443.9	&	2,468.2	&	11.7	&	13.8	\\
1,741.4	&	1,775.3	&	8.4	&	9.9	\\
524.9	&	494.8	&	4.0	&	2.8	\\
306.9	&	286.3	&	3.3	&	1.6	\\
-284.0	&	-320.8	&	0.5	&	-1.8	\\
-221.5	&	-242.3	&	0.1	&	-1.4	\\
869.5	&	877.0	&	4.7	&	4.9	\\
640.7	&	648.7	&	3.6	&	3.6	\\
84.5	&	60.4	&	2.0	&	0.3	\\
-529.1	&	-564.8	&	-1.0	&	-3.1	\\
47.2	&	35.2	&	1.1	&	0.2	\\
451.1	&	463.6	&	2.3	&	2.6	\\
\end{tabular}
\caption{Shapley value attribution to the variance (the first two columns) and volatility (the last two columns) games using Monte Carlo and anlytic expressions.}\label{varianceTab}
\end{table}

\newpage

\section{\label{derSec}P\&L attribution of a derivatives portfolio}

In this section we present the Shapley value approach to the problem of risk (and daily P\&L) attribution of a portfolio of derivatives to a set of possibly non-orthogonal risk factors. The challenge here is to distribute fairly the portion of portfolio P\&L that could in principle be attributed to several risk factors such as delta and vega.

\subsection{\label{sabrSec}SABR model}

In order to  motivate the discussion, we first consider a concrete example, namely a single European option on an asset whose volatility follows the SABR model \cite{HKLW02}. The dynamics of the SABR model is specified in terms of two state variables: forward $F_t$ and instantaneous volatility parameter $\sigma_t$, and is given by:
\begin{equation}
\begin{split}
dF_t&=\sigma_t C(F_t)dW_t,\\
d\sigma_t&=\alpha\sigma_t dZ_t,
\end{split}
\end{equation}
where $W_t$ and $Z_t$ are two Brownian motions with
\begin{equation}
dW_t dZ_t=\rho dt.
\end{equation}
The function $C(x)$ determines the backbone of the volatility smile, and is often assumed to be of the form $C(x)=x^\beta$, where $\beta\leq 1$\footnote{In order to handle negative or close to zero forward rates in interest rate markets, one chooses $C(x)=(x+\theta)^\beta$, with $\theta>0$.}.

In practice, the SABR model is used via an asymptotic approximation in combination with the standard Black-Scholes model. Namely, let $B$ denote the Black-Scholes pricing function in the normal model, i.e.
\begin{equation}
B(\tau,F,K,\sigma)=
\begin{cases}
\sigma\sqrt{\tau}\big(d_+ N(d_+)+N'(d_+)\big),\qquad \text{ for a call option,}\\
\sigma\sqrt{\tau}\big(d_- N(d_-)+N'(d_-)\big),\qquad \text{ for a put option,}
\end{cases}
\end{equation}
where $N(x)$ denotes the cumulative normal distribution, and where
\begin{equation}
d_\pm=\pm\;\frac{F-K}{\sigma\sqrt{\tau}}\;.
\end{equation}
Then the time $t$ value $V_t$ of an option struck at $K$ with expiration $T$ is given by
\begin{equation}
V_t=B(T-t,F_t,K,\sigma^{\mathrm{imp}}(T-t,F_t,K,\sigma_t)).
\end{equation}
Here $\sigma^{\mathrm{imp}}$ is an explicit approximation to the (analytically unknown expression for the) normal volatility implied by the SABR model, see \cite{HKLW02}, \cite{HKLW16} for details. For simplicity, we have disregarded discounting.

This valuation formula is the basis for calculating various sensitivity measures of the option value used to measure its exposure to the factors driving the model. Key among them are the delta
\begin{equation}
\Delta=\frac{\d B}{\d F}+\frac{\d B}{\d\sigma}\frac{\d\sigma^{\mathrm{imp}}}{\d F},
\end{equation}
and vega
\begin{equation}
\Lambda=\frac{\d B}{\d\sigma}\frac{\d \sigma^{\mathrm{imp}}}{\d\sigma}\,,
\end{equation}
calculated as the sensitivities to the ``naive'' scenarios $F_t\to F_t+dF_t,\;\sigma_t\to\sigma_t$, and $F_t\to F_t,\;\sigma_t\to\sigma_t+d\sigma_t$, respectively. It was argued in \cite{B06}, \cite{HL19} that these scenarios are inadequate if $\rho\neq 0$, which may lead in practice to mishedging of the portfolio.

Instead, we proceed as follows. We decompose the Brownian motion $Z_t$ into $W_t$ and a Brownian motion $W^\perp_t$, independent of $W_t$: $Z_t=\rho W_t+\sqrt{1-\rho^2}\,W^\perp_t$. Then, under the forward scenario $F_t\to F_t+dF_t$, the volatility parameter $\sigma_t$ is bound to move as follows: $\sigma_t\to\sigma_t+\rho\alpha/C(F_t)\,dF_t$. Denoting by $d\sigma^\perp_t=\alpha\sigma_t dW_t^\perp$ the component of $d\sigma_t$ uncorrelated with $dF_t$, we obtain from Ito's lemma:
\begin{equation*}
\begin{split}
d\sigma^{\mathrm{imp}}_t&=-\frac{\d \sigma^{\mathrm{imp}}}{\d\tau}\,dt+\Big(\frac{\d \sigma^{\mathrm{imp}}}{\d F}+\frac{\d \sigma^{\mathrm{imp}}}{\d \sigma}\frac{\rho\alpha}{C(F_t)}\Big)dF_t+\frac{\d \sigma^{\mathrm{imp}}}{\d\sigma}\,d\sigma^\perp_t\\
&\quad+\frac12\,\sigma_t^2\Big(C(F_t)^2\,\frac{\d^2 \sigma^{\mathrm{imp}}}{\d^2 F}+2\rho C(F_t)\,\frac{\d^2 \sigma^{\mathrm{imp}}}{\d F\d\sigma}+\alpha^2\,\frac{\d^2 \sigma^{\mathrm{imp}}}{\d^2\sigma}\Big)dt,
\end{split}
\end{equation*}
where $\tau=T-t$. This leads the to the following risk decomposition for the P\&L:
\begin{equation}\label{riskDec}
dV_t=\Big(-\Theta_t+\frac12\,\sigma_t^2\big(C(F_t)^2\Gamma_t+2C(F_t)\mathrm{VA}_t+\alpha^2\mathrm{VO}_t\big)\Big)dt+\Delta^B_t dF_t+\Lambda_t d\sigma^\perp_t,
\end{equation}
where
\begin{equation}\label{bartDel}
\Delta^B=\Delta+\rho\,\frac{\alpha}{C(F)}\frac{\d B}{\d\sigma}\frac{\d\sigma^{\mathrm{imp}}}{\d\sigma}
\end{equation}
is the modified (Bartlett) delta. The other greeks are defined as follows:
\begin{equation}
\Theta=\frac{\d B}{\d\tau}+\frac{\d B}{\d\sigma}\frac{\d\sigma^{\mathrm{imp}}}{\d\tau}
\end{equation}
is the time decay,
\begin{equation}
\Gamma=\frac{\d^2 B}{\d^2 F}+\frac{\d B}{\d\sigma}\frac{\d^2\sigma^{\mathrm{imp}}}{\d F^2}
\end{equation}
is the gamma,
\begin{equation}
\mathrm{VA}=\frac{\d^2 B}{\d F\d\sigma}+\frac{\d B}{\d\sigma}\frac{\d^2\sigma^{\mathrm{imp}}}{\d F\d\sigma}
\end{equation}
is the  vanna, and
\begin{equation}
\mathrm{VO}=\frac{\d^2 B}{\d^2 \sigma}+\frac{\d B}{\d\sigma}\frac{\d^2\sigma^{\mathrm{imp}}}{\d \sigma^2}
\end{equation}
is the volga.

Formula \eqref{riskDec} decomposes the option P\&L in terms of independent risk factors $dF$ and $d\sigma^\perp$, time decay, and second order greeks. It was shown in \cite{B06} and \cite{HL19} that this approach to calculating delta and vega leads to more robust hedges than those obtained by means of the traditionally calculated sensitivities.

Alternatively, we can represent $W_t$ in terms of $Z_t$ and its independent complement $Z^\perp_t$ as $W_t=\rho Z_t+\sqrt{1-\rho^2}\,dZ^\perp_t$, which yields
\begin{equation}\label{altRiskDec}
dV_t=\Big(-\Theta_t+\frac12\,\sigma_t^2\big(C(F_t)^2\Gamma_t+2C(F_t)\mathrm{VA}_t+\alpha^2\mathrm{VO}_t\big)\Big)dt+\Delta_t dF^\perp_t+\Lambda^B_t d\sigma_t.
\end{equation}
Here, 
\begin{equation}
\Lambda^B=\Lambda+\rho\,\frac{C(F)}{\alpha}\Big(\frac{\d B}{\d\sigma}\frac{\d\sigma^{\mathrm{imp}}}{\d F}+\frac{\d B}{\d F}\Big)
\end{equation}
is the modified (Bartlett) vega. Formula \eqref{altRiskDec} is a decomposition of the option's risk in terms of an alternative basis of independent risk factors, namely $dF^\perp$ and $d\sigma$. Note that theta and the second order greeks enter both decompositions in the same manner.

These two forms of risk decomposition show that part of the option's volatility sensitivity can be viewed either as a component of its delta or of its vega:
\begin{equation*}
\begin{split}
dM_t&\equiv \Delta^B_t dF_t+\Lambda_t d\sigma^\perp_t\\
&=\Delta^B_t dF_t+\Lambda^B_t d\sigma_t-\big((\Delta^B_t-\Delta_t)dF_t+(\Lambda^B_t-\Lambda_t)d\sigma_t\big)\\
&=\Delta_t dF^\perp_t+\Lambda^B_t d\sigma_t.
\end{split}
\end{equation*}
In the case of $\rho\neq 0$, there is no obvious way of attributing the P\&L to the delta risk and the vega risk\footnote{A convenient but arbitrary choice is to attribute the overlapping portion of the P\&L to delta.}. We propose that the ``fair attribution'' given by Shapley value is a natural approach.

Notice that the Shapley value of a game with two players is given by  
\begin{equation}
\begin{split}
\mathrm{Sh}_1(v)&=\frac12\,\big(v(\{1,2\})+v(\{1\})-v(\{2\})\big),\\
\mathrm{Sh}_2(v)&=\frac12\,\big(v(\{1,2\})+v(\{2\})-v(\{1\})\big).
\end{split}
\end{equation}
Applying these formulas to $dM$, the martingale part of $dV$, yields
\begin{equation}
\begin{split}
\mathrm{Sh}_{\Delta}(dM)&=\frac12\,\big(\Delta^B dF+\Delta dF^\perp\big),\\
\mathrm{Sh}_{\Lambda}(dM)&=\frac12\,\big(\Lambda^B d\sigma+\Lambda d\sigma^\perp\big).
\end{split}
\end{equation}
This can alternatively be expressed in the following form:
\begin{equation}
\begin{split}
\mathrm{Sh}_{\Delta}(dM)&=\Delta^B dF-\frac12\,\big((\Delta^B-\Delta)dF+(\Lambda^B-\Lambda)d\sigma\big),\\
\mathrm{Sh}_{\Lambda}(dM)&=\Lambda^B d\sigma-\frac12\,\big((\Delta^B-\Delta)dF+(\Lambda^B-\Lambda)d\sigma\big).
\end{split}
\end{equation}
In other words, Shapley value allocates half of the common risk components to each delta and vega risks. Note that when $\rho=0$,
\begin{equation}
\begin{split}
\mathrm{Sh}_{\Delta}(dM)&=\Delta dF,\\
\mathrm{Sh}_{\Lambda}(dM)&=\Lambda d\sigma,
\end{split}
\end{equation}
as expected.

\subsection{\label{genModSec}General case}

We can now extend this discussion to the general case of an arbitrary number of risk factors. We consider a portfolio of derivatives whose price dynamics is captured by a multifactor diffusion process:
\begin{equation}
\begin{split}
dX_{t,j}&=D_j(t,X_t)dW_{t,j},\\
dW_{t,j} dW_{t,k}&=\rho_{jk}dt.
\end{split}
\end{equation}
The state vector $X_t\in\bR^n$ may consist of both market observable factors such as prices, yields, exchange rates, etc, as well as of unobservable factors such as parameters capturing stochastic volatility, convenience yields, etc. For convenience, we work under the measure under which the pricing process is a martingale. The portfolio value process is given by $V_t=V(t, X_t)$, where $V(x)$ is a valuation function.

Traditional way to calculate the sensitivity of the portfolio value to $X_i$ is via the infinitesimal scenario:
\begin{equation}
\begin{split}
X_i&\to X_i+dX_i,\\
X_j&\to X_j,\text{ for }j\neq i.
\end{split}
\end{equation}
This leads to the usual sensitivities,
\begin{equation}
\Delta_i=\nabla_i V,
\end{equation}
where $\nabla_i$ denotes partial derivative.

Alternatively, and more appropriately, in order to account for the correlation effects among the risk factors, we consider a scenario under which $X_i$ moves by $dX_i$:
\begin{equation}
X_i\to X_i+dX_i,
\end{equation}
accompanied by the induced moves of the remaining state variables:
\begin{equation}
X_j\to X_j+\rho_{ji}\,\frac{D_j(X)}{D_i(X)}\,dX_i+dW^\perp_j.
\end{equation}
Here, each $dW^\perp_j$ is uncorrelated with $dW_i$. Hence the \textit{correlation adjusted sensitivity} (generalized Bartlett's delta) is given by
\begin{equation}
\fD_i=\nabla_i V +\sum_{j\neq i}\rho_{ij}\;\frac{D_j(X)}{D_i(X)}\;\nabla_j V,
\end{equation}
or
\begin{equation}
\fD_i=\Delta_i +\sum_{j\neq i}\rho_{ij}\;\frac{D_j(X)}{D_i(X)}\;\Delta_j.
\end{equation}

We can thus write
\begin{equation}
dV=\dot{V}dt+\frac12\,\var(dX)+\fD_i dX_i+\sum_{j\neq i}\,\Delta_j dX^\perp_j,
\end{equation}
where $\dot{V}$ denotes time derivative. This equation expresses the impact of a move in the state variable $X_i$ on the value of the asset in terms of the correlation adjusted sensitivity to the variable $X_i$, and the classic sensitivities multiplying the moves of the remaining variables with the components explained by the change in $X_i$ removed.

Calculating Shapley value of the martingale part $dM$ of $dV$ is straightforward, and it yields
\begin{equation}\label{genBart}
\mathrm{Sh}_i(dV)=\frac{1}{n}\;\Big(\fD_i dX_i+\sum_{j\neq i}\,\Delta_j dX^\perp_j\Big).
\end{equation}
The variance part of $dV$ can be attributed using equation \eqref{varShap}. As a result we find that the P\&L attribution to the $i$-th factor is given by 
\begin{equation}\label{genBart}
\mathrm{Sh}_i(dV)=\frac{1}{n}\;\Big(\fD_i dX_i+\sum_{j\neq i}\,\Delta_j dX^\perp_j\Big)+\frac12\,\cov(dX_i,dX).
\end{equation}
Notice that, strictly speaking, time decay is also part of the P\&L attribution process. We have excluded it from the discussion, as it is not a risk factor and its attribution is straightforward.

In practice, the infinitesimal quantities $dX_i$ are replaced by the (daily) changes $\delta X_i$, and the equation above takes the form of the following approximation:
\begin{equation}\label{genBart}
\mathrm{Sh}_i(\mathrm{P\&L})\approx\frac{1}{n}\;\Big(\fD_i\, \delta X_i+\sum_{j\neq i}\,\Delta_j\, \delta X^\perp_j\Big)+\frac12\,\cov(\delta X_i,\delta X).
\end{equation}
The difference $\mathrm{P\&L}-\sum_i \mathrm{Sh}_i(\mathrm{P\&L})$ is the unexplained P\&L.

\section{\label{varSec}Attribution of VaR and ES}

In this section we discuss how Shapley value can be used to allocate enterprise risk metrics such as VaR and ES to the risk factors of the portfolio. This can be accomplished regardless of the methodology used for VaR and ES estimation.

\subsection{\label{ellSec}Elliptic distribution of losses}

Value at Risk and Expected Shortfall are risk metrics typically used at the enterprise risk management level. Their primary goal is to determine the amount of cash reserves required to be held against potential portfolio losses at a given confidence level defined as $q\equiv 1-\alpha$ (say, 95\%), where $\alpha$ is the significance level. Among the approaches to VaR and ES, the approach based on a parametric distribution of losses is the most straightforward, albeit not the most commonly used. Its main advantage is the conceptual simplicity and ease of estimation: only a well defined set of parameters requires estimation. On the negative side, this approach implicitly assumes stationarity of portfolio returns and often leads to unrealistic tail dependence.

A large class of examples of relevant parametric distributions is provided by the family of multifactor elliptical distributions \cite{CHS81}. Recall that a random vector $X\in\bR^n$ has an elliptical distribution with parameters $\mu\in\bR^n$ and positive definite $\Sigma\in\matn$, if its density function is of the form $c_n \det(\Sigma)^{-1/2}h\big((x-\mu)^\tT \Sigma^{-1}(x-\mu)\big)$, with a nonnegative function $h$ on $\bR$ and a normalizing constant $c_n$. Assuming that the distribution has moments of order one and two (which we are doing), $\eE(X)=\mu$, and $\cov(X)=\kappa_n\Sigma$, where $\kappa_n$ is a constant. Important examples of an elliptical distribution are the multivariate Gaussian distribution, multivariate Student t distribution, and Laplace distribution.

Let $X$ represent the portfolio's daily P\&L. Under an elliptical distribution, $\mathrm{VaR}_q$ and $\mathrm{ES}_q$ are linear in the standard deviation $\sigma(X)=\sqrt{\var(X)}$,
\begin{equation}
\mathrm{VaR}_q(X)=\mu(X)-A_q\sigma(X),
\end{equation}
and
\begin{equation}
\mathrm{ES}_q(X)=\mu(X)-B_q\sigma(X).
\end{equation}
Here, $A_q$ and $B_q$ are constants depending only on the distribution and confidence level, and $\mu(X)$ is linear in $X$. For example, if $X$ is distributed according to the the multinomial Gaussian distribution $N(\mu,\Sigma)$ and $\sigma(X)=\sqrt{\var(X)}$, then
\begin{equation}
\mathrm{VaR}_q(X)=\mu^\tT X-\Phi^{-1}(q)\sigma(X),
\end{equation}
where $\Phi(z)$ is the cumulative standard normal distribution, and
\begin{equation}
\mathrm{ES}_q(X)=\mu^\tT X-\frac{\varphi(\Phi^{-1}(q))}{1-q}\,\sigma(X),
\end{equation}
where $\varphi(z)=(2\pi)^{-1/2})\exp(-z^2/2)$ is the density of $\Phi(z)$. Analogous explicit expressions exist for the Student t distribution, Laplace distribution, and others.

In order to find the Shapley value of (the nonlinear cooperative games defined by) $\mathrm{VaR}_\alpha$ and $\mathrm{ES}_\alpha$, we note first that the Shapley value of (the linear cooperative game defined by) $\mu$ is given by $\mathrm{Sh}_i(X)=\mu(X_i)$. Using the additive property of Shapley value, and the approximation \eqref{apprSh}, we find that
\begin{equation}
\mathrm{Sh}_i(\mathrm{VaR}_q)\approx\mu(X_i)-A_q\rho(X_i,X)\sigma(X_i),
\end{equation}
and
\begin{equation}
\mathrm{Sh}_i(\mathrm{ES}_q)\approx\mu(X_i)-B_q\rho(X_i,X)\sigma(X_i).
\end{equation}

\subsection{\label{hvarSec}Historical bootstrapping}

Historical bootstrapping (and variations thereof) is the default methodology for VaR and ES estimation. In this approach VaR is estimated as the relevant quantile of the empirical distribution of the (daily) P\&L of a portfolio, calculated over a specified lookback period of historical data. Expected shortfall is estimated as the arithmetic average of the losses exceeding the estimated VaR level. This approach has the advantage that it is blind to the stationarity and tail properties of the portfolio return distribution. 

For example, in order to find the 95\% VaR with a two year lookback period, we (i) calculate the P\&L over the past 500 business days, (ii) rank it in the ascending order $PL_{1}\leq\ldots\leq PL_{500}$ (the largest loss first) and (iii) set $\mathrm{VaR}_{95\%}=PL_{25}$. The estimated expected shortfall is $\mathrm{ES}_{95\%}=\tfrac{1}{25}\sum_{j=1}^{25} PL_j$.

The so estimated VaR and ES can be easily attributed to the predefined risk factors using Shapley value. To this end, for each chosen risk factor $j$ we select a subset $X_j$ of the set of assets $X$ that represents this risk factor. Note that the subportfolios $X_j$ representing different risk factors are not required to be mutually disjoint. This allows us to assign a particular asset to several risk groups. For example, in case of a portfolio of equities, these subsets can represent the stock's industry, size, liquidity, etc.

Using the paradigm of historical bootstrapping, we can estimate the values $\mathrm{VaR}_q(S)$ and $\mathrm{ES}_q(S)$ of a coalition $S$ of the factors defined by the subportfolios $X_j$. To this end we define the following characteristic functions $\mathrm{VaR}^h$ and $\mathrm{ES}^h:\,2^\mathscr{N}\to\bR$:
\begin{equation}
\mathrm{VaR}^h(S)=\text{historical VaR estimated off the series }X_{t,S}=\sum_{j\in S}X_{t,j},\,t=1,\ldots,500,
\end{equation}
and
\begin{equation}
\mathrm{ES}^h(S)=\text{historical ES estimated off the series }X_{t,S}=\sum_{j\in S}X_{t,j},\,t=1,\ldots,500.
\end{equation}
This allows us to allocate the portfolio VaR or ES (estimated via historical bootstrapping) to each of the factors: (i) through an exact calculation, if feasible, or (ii) via Monte Carlo simulations, otherwise (in case of a large number of factors).

\subsection{\label{numVarSec}Numerical simulations}

Table \ref{varTab} a numerical test of the concepts discussed in this section. To this end, we reuse the simulated data of Section \ref{numCovSec}. This data is meant to simulate daily P\&L of a portfolio of 25 assets observed over the period of two years. Its historical VaR and ES at the 95\% level of confidence are given by -302.4 and -358.3, respectively, and the corresponding values estimated from multifactor Gaussian model are -295.2 and -370.2. Not surprisingly they are close to each other as they represent finite sample estimates of the same quantities.

The meaning of the columns in Table \ref{varTab} is as follows:
\begin{itemize}
\item[1.]{HV represents $\widehat{\mathrm{Sh}}_i(\mathrm{VaR}^h)$,}
\item[2.]{GV represents $A_q\rho(X_i,X)\sigma(X_i)$,}
\item[3.]{HE represents $\widehat{\mathrm{Sh}}_i(\mathrm{ES}^h)$,}
\item[4.]{GE represents  $B_q\rho(X_i,X)\sigma(X_i)$,}
\end{itemize}
for $i=1,\ldots,25$. As before, for the Monte Carlo computations of the Shapley value (columns one and three), we used used 100,000 simulations.
 
\begin{table}[H]
\centering
\begin{tabular}{l | l || l | l}
HV& GV & HE	& GE\\
\hline\hline
-10.5	&	-10.9	&	-12.5	&	-13.7	\\
-22.0	&	-16.8	&	-25.9	&	-21.1	\\
-33.7	&	-30.8	&	-39.6	&	-38.6	\\
-30.8	&	-37.1	&	-39.7	&	-46.5	\\
-14.3	&	-11.3	&	-15.7	&	-14.2	\\
-12.8	&	-12.0	&	-15.5	&	-15.1	\\
-19.3	&	-27.8	&	-27.3	&	-34.9	\\
-27.9	&	-39.5	&	-33.2	&	-49.5	\\
-14.9	&	-6.8	&	-15.4	&	-8.5	\\
-24.7	&	-31.9	&	-31.3	&	-40.0	\\
-10.8	&	-8.5	&	-11.6	&	-10.7	\\
-5.5	&	-1.3	&	-6.6	&	-1.7	\\
-3.5	&	-6.3	&	-4.4	&	-7.9	\\
-19.4	&	-22.6	&	-24.5	&	-28.4	\\
-18.1	&	-16.3	&	-24.6	&	-20.4	\\
-8.0	&	-4.5	&	-9.7	&	-5.7	\\
-4.1	&	-2.6	&	-6.7	&	-3.3	\\
0.0	&	2.9	&	0.9	&	3.7	\\
-0.9	&	2.2	&	-0.7	&	2.8	\\
-8.3	&	-8.0	&	-9.7	&	-10.1	\\
-3.3	&	-5.9	&	-3.0	&	-7.5	\\
-3.3	&	-0.6	&	-3.9	&	-0.7	\\
0.7	&	5.2	&	1.2	&	6.5	\\
-4.4	&	-0.3	&	-5.2	&	-0.4	\\
-3.3	&	-4.2	&	-3.5	&	-5.3	\\
\end{tabular}
\caption{Shapley value attribution of VaR (the first two columns) and ES (the last two columns) using historical and Gaussian estimates}\label{varTab}
\end{table}

\end{document}